\documentclass[letterpaper, 10 pt, conference]{ieeeconf}  

\IEEEoverridecommandlockouts                              

\usepackage[utf8]{inputenc}
\usepackage{url}
\usepackage{hyperref}
\usepackage{graphicx}
\usepackage{amsmath,amssymb,amsfonts}
\usepackage{cite}
\usepackage{algorithm}
\usepackage{algpseudocode}
\usepackage{graphicx,psfrag,epsf}
\usepackage{xcolor}
\usepackage[caption=false]{subfig}
\usepackage{tabularx}
\usepackage{float}
\usepackage{multirow}
\usepackage[section]{placeins}
\usepackage{textcomp}
\pdfoutput=1
\usepackage{graphicx}
\usepackage{array}
\usepackage{amsmath}
\usepackage{graphicx,psfrag,epsf}
\usepackage{enumerate}
\usepackage{url} 
\usepackage{hyperref}
\usepackage{graphicx} 
\usepackage{amsthm,amssymb}
\usepackage{bm}
\usepackage{hyperref}
\usepackage{dsfont}
\usepackage[caption=false]{subfig}
\usepackage{tabularx}
\usepackage{float}
\usepackage{algorithm}
\usepackage{algpseudocode}
\usepackage{multirow}
\usepackage{todonotes}

\title{\LARGE \bf
Operator-Based Detecting, Learning, and Stabilizing Unstable Periodic Orbits of Chaotic Attractors
}

\author{Ali Tavasoli$^{1}$ and Heman Shakeri$^{2}$
\thanks{$^{1}$Ali Tavasoli is with the School of Data Science, University of Virginia, Charlottesville, VA, USA
        {\tt\small at9kf@virginia.edu}}%
\thanks{$^{2}$Heman Shakeri is with the School of Data Science, University of Virginia, Charlottesville, VA, USA
        {\tt\small hs9hd@virginia.edu}}%
}

\begin{document}

\maketitle
\thispagestyle{empty}
\pagestyle{empty}

\begin{abstract}

This paper examines the use of operator-theoretic approaches to the analysis of chaotic systems through the lens of their unstable periodic orbits (UPOs). Our approach involves three data-driven steps for detecting, identifying, and stabilizing UPOs. We demonstrate the use of kernel integral operators within delay coordinates as an innovative method for UPO detection. 
For identifying the dynamic behavior associated with each individual UPO, we utilize the Koopman operator to present the dynamics as linear equations in the space of Koopman eigenfunctions. 
This allows for characterizing the chaotic attractor by investigating its principal dynamical modes across varying UPOs. 
We extend this methodology into an interpretable machine learning framework aimed at stabilizing strange attractors on their UPOs. To illustrate the efficacy of our approach, we apply it to the Lorenz attractor as a case study.

\end{abstract}

\section{INTRODUCTION}

Over the past few decades, machine learning and deep learning techniques have dramatically transformed various fields of science and engineering \cite{Brunton_Kutz_2019}. 
As dynamic systems evolve, nonlinear interactions can result in chaotic and mixing behaviors, giving rise to uncorrelated spaces. This complexity is not limited by the form of interactions; even the simple nonlinear systems can exhibit complex behavior, as seen when a quadratic term leads to chaotic dynamics in the R\H{o}ssler model \cite{Sprott_book}. This underscores the concept of natural subspaces for embedding complex non-linear systems within finite, closed spaces, thus bypassing the need for implicit function spaces in machine learning tools \cite{Coifman2008stochastic, Giannakis2013NLSA}. 

Considerable progress has been made in fulfilling this requirement in recent years after revisiting the \textit{Koopman operator} through efficient data-driven approaches \cite{Mezic2005Spectral, Williams2015EDMD, Mezic2020Spectral, Giannakis2019Spectral, Giannakis2019Delay, Colbrook2023resDMD}. 
The Koopman operator constructs linear models in the observable space, bypassing the need for nonlinear models in the state space. Koopman eigenfunctions offer an optimal coordinate set for reconstructing the dynamics, and the reduced-order models based on these rich dynamical elements encapsulate maximum information about intricate nonlinear interactions in interconnected systems.
Studying the projected dynamics into each Koopman eigenspace will unveil a portion of the overall dynamics by highlighting independent nonlinear features specific to that eigenspace. 
By collating all these projected images, we can construct a comprehensive view of the large-scale dynamics in a flexible manner. This global linear representation is analogous to generalizing the Hartman--Grobman theorem that encompasses the entire basin of equilibrium points or periodic orbits \cite{Brunton2016Koopman}. Alternatively, it could be considered a counterpart to the principal modes of SVD approaches in linear theory \cite{Giannakis2013NLSA}.

Accordingly, the Koopman operator provides a theoretical framework for the spectral analysis and mode decomposition of nonlinear systems \cite{Mezic2005Spectral, Rowley2009Spectral}. 
Unlike traditional methods such as Fourier analysis and POD, which are commonly used to decompose signals into their frequency contents \cite{Holmes1996}, the spectra of the Koopman operator facilitate the decomposition of the dynamics on an optimal nonlinear basis \cite{Brunton2021Fourier} and  reveal spatial patterns corresponding to each temporal mode. 
Utilizing this spatiotemporal mode decomposition approach in nonlinear systems  allows for investigation of the roles and interactions of constituents  within different temporal modes. Hence, it facilitates the discovery of mechanisms and phenomena governing the operation of the overall system.

More specifically, Koopman eigenfunctions recover coherent patterns with associated timescales \cite{Giannakis2019Spectral}, offering a unique setting for studying complex systems. These systems can exhibit diverse dynamical properties interacting across varying scales with a broad spectrum range. Therefore, this spectral analysis provides deep insights into the dynamics, such as invariant sets and partitions, basins of attractions for limit cycles, fixed points, and stability \cite{Mezic2020Book}. 


One way to characterize chaotic attractors is examining the unstable periodic orbits (UPOs) that are embedded densely within them \cite{Franceschini1993_Characterization}. UPOs make up the skeleton
of a chaotic attractor, and an orbit on the attractor is the closure of the set of UPOs.
The trajectory can be thought of as hopping from one UPO to the next \cite{Bradley2002recurrence}. 
The set of UPOs in an attractor is a dynamical invariant; their number, distribution, and properties unfold the structure of chaotic orbits, and they can be used to calculate other invariants, such as fractal dimension and topological entropy \cite{Franceschini1993_Characterization, Bradley2002recurrence}. 

Though UPOs in time series data may be identified by watching for close returns on a plane of section, this procedure is quite time-consuming due to the involvement of an ensemble of nearest-neighbor searches to reduce noise \cite{Bradley2002recurrence}. 
One can accelerate matters somewhat by using estimates of the local dynamics \cite{Ott1990OGY, Ott1996detecting, Ott1997extracting}, but the computational complexity is largely inescapable \cite{Bradley2002recurrence}.

The detection of UPOs leads to a natural way for controlling chaotic systems towards regular dynamics while retaining the original dynamics features. 
Indeed, by making only small control perturbations, chaotic dynamics can be tamed by exploiting the UPOs that exist without control, thereby avoiding creation of new orbits with different properties  ones \cite{Ott1990OGY, Ott1994observing, Ott1995controlling, Ott1996detecting, Ott1997extracting, Ott_Book}. As the uncontrolled orbit wanders ergodically over the attractor, it will eventually approach every (selected) UPO. When this happens, a small kick is sufficient to align the orbit with the UPO.

 Although requiring the dynamical equations and detecting UPOs in high-dimensional data spaces pose substantial challenges \cite{Ott1990OGY, Ott1994observing, Ott1995controlling, Ott1996detecting, Ott1997extracting}, as this paper demonstrates, the operator-theoretic framework provides a practical data-driven approach for detecting, identifying, and stabilizing UPOs.

Our approach to detecting UPOs is based on examining the kernel integral operators in delay coordinates for the given data set. Recently studies have demonstrated the application of kernel operators for discovering various inherent dynamical properties in data-driven settings \cite{Giannakis2015Kernel, Williams2015Kernel, Berry2016Variable, Klus2019kernel,   Colbrook2023resDMD}. Further works on kernel operators in delay coordinates  \cite{Giannakis2019Spectral, Giannakis2019Delay} and reproducing kernel Hilbert spaces (RKHS) \cite{Giannakis2021RKHS, Klus2020RKHS, Fjii2019GraphDMD, Hamzi2021partI} have established a solid theoretical foundation for recovering Koopman invariant subspaces and devising advanced machine learning techniques for complex nonlinear dynamics.

After detecting UPOs using the kernel operator, we use the Koopman mode decomposition to identify the evolving dynamics on them. Hence, each detected UPO can be represented through principle modes that evolve linearly in the Koopman eigenfunctions space. While this interpretable learning of UPOs might be used to explore complex strange attractors, we demonstrate its further use for chaos control in next step. This involves two methods based on tracking the identified UPO's dynamics and small control perturbation based on the UPO's Koopman eigenfunctions.

The rest of the paper is organized as follows. The next section outlines our approach based on the Koopman and kernel operators. Section \ref{sec:method} demonstrates the use of this approach for detecting, identifying, and controlling UPOs. Section \ref{sec:results} presents the results of applying this approach to the Lorenz attractor. Section \ref{sec:cnclsn} is devoted to concluding remarks.

\section{Koopman operator and generator}\label{sec:Koopman}
The Koopman approach considers the available data set as observables of a dynamical system with state space $\mathcal X$ and flow map $\Phi^t : \mathcal X \mapsto \mathcal X$, where $t\in\mathbb R$ denotes time. The system's state at time $t$ is given by $x_t=\Phi^t(x)$, where $x\in \mathcal X$. Signal recorded at $d$ sensors are considered as an observation function $F:\mathcal X\mapsto\mathbb R^d$. 
The dynamical system $(\mathcal X , \Phi^t)$ possesses ergodic measures; hence there exists a probability measure $\mu$ on $\mathcal X$, invariant under the flow map $\Phi^t$, such that for every integrable function $f : \mathcal X\mapsto\mathbb C$, the time average $\bar f$ of $f$ converges to the expectation value $\bar f = \int_\mathcal X f d\mu$. 
Associated with the triplet $(\mathcal X , \Phi^t , \mu)$, we consider a Hilbert space $\mathcal H = L^2(\mathcal X , \mu)$ of square-integrable observables with respect to $\mu$. 

The group of unitary Koopman operators $U^t:\mathcal H\mapsto\mathcal H$ governs the evolution of observables under $\Phi^t$. That means given $f\in\mathcal H$, $g=U^tf$ is defined as the observable satisfying $g(x)=f(\Phi^t(x))$ for $x\in \mathcal X$. 
An observable $\psi_j\in\mathcal H$ is a Koopman eigenfunction if it satisfies the eigenvalue equation
\begin{equation}\label{eq:eigen}
    \begin{gathered}
        U^t\psi_j=e^{i\omega_jt}\psi_j
    \end{gathered}
\end{equation}
for all $t \in \mathbb R$. The eigenfrequency $\omega_j$ is a real-valued frequency associated with the eigenfunction $\psi_j$. Therefore, in measure-preserving dynamical systems, the Koopman eigenvalues remain on the unit circle in the complex plane, and the corresponding eigenfunctions evolve periodically under the dynamics. This is the key to the predictability of coherent patterns of dynamics. 
The Koopman eigenvalues and eigenfunctions appear as complex-conjugate pairs, and the Koopman eigenfunctions that correspond to different eigenfrequencies are orthogonal in the Hilbert space $\mathcal H$.  

For every continuous flow $\Phi^t$, the family of operators $U^t$ has a generator $V$, which is a skew-adjoint operator, defined as 
\begin{equation}\label{eq:Generator}
\begin{gathered}
Vf:=\lim_{t\mapsto 0}\frac{1}{t}(U^tf-f), \ \ \ f\in D(V)\subset L^2(\mathcal X,\mu)
\end{gathered}
\end{equation}
Operators $U^t$ and $V$ share the same eigenfunctions,
\begin{equation}
    \begin{gathered}
    V\psi_j=i\omega_j\psi_j
    \end{gathered}
\end{equation}
For ergodic systems, all eigenvalues of $V$ are simple.


\section{DYNAMICS ADAPTED KERNEL INTEGRAL OPERATORS}\label{sec:data-driven}
\subsection{Kernel operators and delay coordinates}\label{sec:Kernel}
 
Given a a collection of $N$ samples $F(x_1),...,F(x_N)$, organized in a time-ordered manner, where each $F(x_i)\in \mathbb{R}^d$. The value of $x_n$ is determined by the function $\Phi^{n\Delta t}(x_0)$, where $\Delta t$ is the interval at which the data is sampled. A delay coordinate
map is constructed from $F$ by embedding $\mathcal X$ in a manifold in $\mathbb R^{Qd}$ as
\begin{equation}
    \begin{gathered}
    F_Q(x) = (F(x),F(\Phi^{-\Delta t}x), \cdots , F(\Phi^{-(Q-1)\Delta t}(x)))
    \end{gathered}
\end{equation}
where $Q$ (an integer) is the number of delays. Next, a kernel function $k_Q : \mathcal X\times \mathcal X\mapsto \mathbb R_+$ is defined to measure the similarity of points in $\mathcal X$ based on the observation function $F_Q$. In this work, we use the radial Gaussian kernel with variable bandwidth defined as:
\begin{equation}
\begin{gathered}
k_Q(x,x')=\exp({-\frac{\parallel F_Q(x)-F_Q(x')\parallel^2}{\epsilon}})
\end{gathered}
\end{equation}
where $\epsilon$ is a positive bandwidth parameter that can vary based on the available data density in $\mathcal X$. In this work, we use a class of variable bandwidth kernels, also known as self-tuning kernels, introduced in \cite{Berry2016Variable}. 

Associated with the square-integrable kernel $k_Q$ is a compact integral operator,
\begin{equation}
    \begin{gathered}
    K_Qf(x) := \int_\mathcal Xk_Q(x,y)f(y)d\mu(y)
    \end{gathered}
\end{equation}
Then a Markov kernel is constructed by normalizing $K_Q$, 
\begin{equation}
    \begin{gathered}
    P_Qf:=\frac{\tilde K_Qf}{\tilde K_Q1_\mathcal X}
    \end{gathered}
\end{equation}
where $\tilde Kf=K_Q(\frac{f}{K_Q1_\mathcal X})$ and $1_\mathcal X$ is the indicator function for the set $\mathcal X$. The matrix $P$ is a Markov matrix with real eigenvalues ordered as  $1=\lambda_1>\lambda_2\geq\lambda_3\geq...$, and real eigenvectors $\varphi_j$ that are mutually orthogonal in $\mathbb R^N$. Note that the first eigenvector corresponding to $\lambda_1=1$ is the constant eigenvector $\varphi_1=(1,...,1)^T\in\mathbb R^N$.

As $Q$ approaches infinity, the kernel integral operator $P$ commutes with $U^t$ \cite{Giannakis2019Delay}. Commuting operators have a common eigenspace, which allows us to compute eigenfunctions of $U^t$ through expansions in the basis obtained from $P$. Compactness of $P$ makes a wide variety of data-driven tools available for approximation of integral operators.
\subsection{Galerkin approximation of Koopman spectra}\label{sec:Galerkin}
To solve the numerically ill-posed Koopman eigenvalue problem, we replace finding the eigenvalues of $V$ with approximating them through a regularized operator $L_\theta$ with a small amount $\theta$ of judiciously added diffusion for regularization \cite{Giannakis2019Spectral, Giannakis2019Delay}. Eigenfunctions of the Koopman operator are sought in a Sobolev subspace $\mathcal H^2\subset\mathcal H$ \cite{Giannakis2019Delay}, on which $V$ is a bounded operator. 
Based on the standard Galerkin approach, the regularized Koopman eigenvalue problem is to find $\gamma\in\mathbb C$ and $z\in\mathcal H^2$, such that for every $f\in\mathcal H$ the following weak formulation holds:
\begin{equation}\label{eq:Galerkin}
    \begin{gathered}
    \left<z,L_\theta f\right>=\gamma\left<f,z\right>, \ \ \ L_\theta=V-\theta\Delta
    \end{gathered}
\end{equation}
where the operator $\Delta$ is defined based on the eigenfunctions of the Laplace-Beltrami operator \cite{Giannakis2019Spectral, Giannakis2019Delay}. In \eqref{eq:Galerkin}, $\gamma$ and $z$ are weak eigenvalues and eigenfunctions of $L_\theta$, and $f$ is a test function. To solve the Galerkin problem \eqref{eq:Galerkin}, we use the eigenfunctions $\varphi_j$ of the kernel operator $P$ as a basis.
While the regularized operator $L_\theta$ and the generator $V$ share the same eigenfunctions, the eigenvalues of $L_\theta$ are parameterized as $\gamma_\theta=i\omega-\theta\eta$ where $i\omega$ is an eigenvalue of $V$ and $\eta$ an eigenvalue of $\Delta$. Next, we order the eigenfunctions by increasing Dirichlet energy values, denoted by $E(f)=\frac{\left<f,\Delta f\right>}{\parallel f\parallel^2}$.

\subsection{Nystrom extension of Koopman eigenfunctions} \label{sec:Nystrom}

To evaluate the Koopman eigenfunctions at out-of-sample points, we first apply the Nystrom approach to extend the eigenfunctions $\varphi_k$ of the Markov kernel operator \cite{Coifman2006Nystrom}. Then, we use the results of the Galerkin approach for out-of-sample evaluation of Koopman eigenfunctions.

Assuming that the Markov integral operator $P$ described in Section \ref{sec:Kernel} has the kernel $p:\mathcal X\times \mathcal X\mapsto\mathbb R_+$, and the data $X_s = \{x_1, x_2,..., x_N\}$, sampled from the manifold $\mathcal X$, as $N$ landmark points at which the eigenfunction $\varphi$ is sampled. Consider $\hat\varphi(x)$ as an approximation to the true $\varphi(x)$, and the corresponding eigenvalue $\hat\lambda_i$. Then for an unsampled point $x$, we have
\begin{equation}
    \begin{gathered}
    \hat\varphi_i(x)=\frac{1}{\hat\lambda_i}\sum_{j=1}^Np(x,x_j)\hat\varphi_i(x_j)
    \end{gathered}
\end{equation}

\section{Extracting, identifying, and controlling UPOs}\label{sec:method}

\subsection{Extracting UPOs}
\begin{figure}
      \centering
      \includegraphics[width=1\columnwidth]{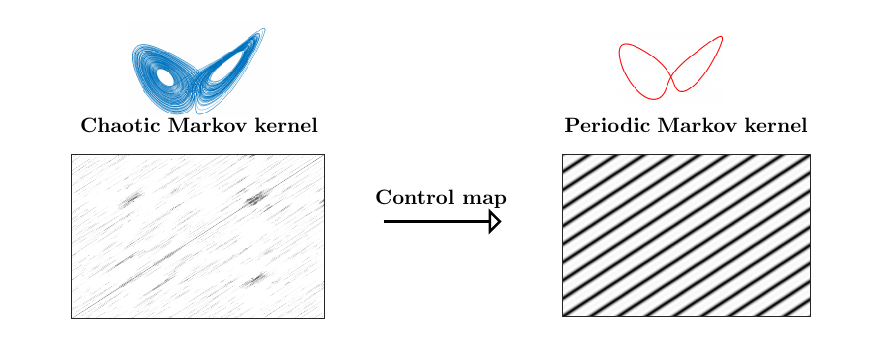}
      \caption{Mapping the Markov operator between chaotic and periodic patterns by active control.}
      \label{fig:GA}
\end{figure}

In Figure \ref{fig:GA}, we show the main idea of mapping between chaotic and periodic Markov kernel operators. The Markov kernel $P_Q$ indicates a regular pattern for periodic orbits, where recurrent system states are encoded as high-probability transitions (black patches). The chaotic Markov kernel exhibits an irregular pattern with transition probabilities scattered sporadically. Thus, transitions between different chaotic states are random, resulting in unpredictable and mixing dynamics. 

To implement our approach, we leverage the fact that chaotic attractors are dense with UPOs. When a trajectory nears a UPO's stable manifold, it evolves (almost) periodically according to the UPO properties \cite{Ott1990OGY}. 
By zooming into parts of $P_Q$ corresponding to periodic time intervals, we recover the regular pattern in Figure \ref{fig:GA}. 
Our goal is to discover the regular pattern patches around the diagonal of $P_Q$ (see Figure \ref{fig:kernel}). This sets out the application of modern machine learning techniques to discover UPOs via kernel operators in delay coordinates, and serves as an alternative to traditional approaches reliant on low-order maps or underlying model equations \cite{Ott_Book}.  

\subsection{Identification and control}

We formulate the system's dynamics on a detected UPO by computing the Koopman eigenfunctions for the trajectory on that UPO. We consider the system state on a UPO as an observable $\boldsymbol x_d\in\mathcal D$ in the space of Koopman eigenfunctions that is invariant under $U^t$, and is the closure of the span of ${\psi_j}$.  
Every $f \in \mathcal D$ can be decomposed as $f=\sum_j\hat f_j\psi_j$, where $\hat f_j=\left<f,\psi_j\right>_\mathcal H$ is the inner product in $\mathcal H$ \cite{Giannakis2019Spectral}.
Moreover, the dynamical evolution of $f$ can be computed in a closed form via
\begin{equation}\label{eq:evolution}
    \begin{gathered}
    U^tf = \sum_j\hat f_je^{i\omega_j}\psi_j
    \end{gathered}
\end{equation} 
This allows us to represent the system trajectory on a UPO as the following linear output dynamics,
\begin{equation}\label{eq:x_d}
    \begin{gathered}
        \dot{\boldsymbol\psi}=\Lambda\boldsymbol\psi, \hspace{6mm}
            \boldsymbol x_d=C\boldsymbol\psi
    \end{gathered}
\end{equation}
where $\boldsymbol\psi=[\psi_1,\dots,\psi_N]^T$ is the vector of $N$ Koopman eigenfunctions in the invariant subspace, $\Lambda_{N\times N}$ is a diagonal matrix with the generator eigenvalues on the main diagonal, and $C_{d\times N}$ is the regression matrix projecting the state space variables to the Koopman invariant subspace \cite{tavasoli2023characterizing}. 

Thus, we collect data over the UPO and extract the Koopman eigenfunctions $\psi_j$ using the Galerkin formulation \eqref{eq:Galerkin}. The extracted dynamics may serve as a reference for closed-loop control. We consider two control architectures. The first control law tracks the UPO-induced trajectory $\boldsymbol x_d(t)$ by simply computing the error $\boldsymbol e(t)=\boldsymbol x(t)-\boldsymbol x_d(t)$, where $\boldsymbol x(t)$ represents the actual system trajectory. The second method leverages the ergodic property of chaotic attractors and employs a small-perturbation policy, activating the control only if the system trajectory remains within a sufficiently small distance of the UPO's stable manifold \cite{Ott1990OGY}. To test this condition, the proposed algorithm monitors how Koopman eigenfunctions extend to the evolving system trajectory (see Section \ref{sec:LorenzCntrl}). 

\section{Application to Lorenz attractor} \label{sec:results}

The Lorenz equations for fluid convection in a two-dimensional layer heated from below are 
\begin{equation}
    \begin{gathered}
        \dot x = \sigma(y-x) , \ \
        \dot y = -xz+rx-y , \ \
        \dot z = xy-bz
    \end{gathered}
\end{equation}
where $\sigma$, $r$, and $b$ are system parameters. The system state is $\boldsymbol x=[x \ y \ z]^T\in\mathbb R^3$. For the Lorenz-63 model, we set $\sigma=10$, $r=28$, and $b=8/3$. This is a chaotic attractor with the mixing property \cite{Luzzatto2005mixingLorenz}.

\subsection{Detecting UPOs in Lorenz model}
\begin{figure}
    \centering
    \subfloat[\label{fig:kernal_a}]{\includegraphics[width=0.49\columnwidth]{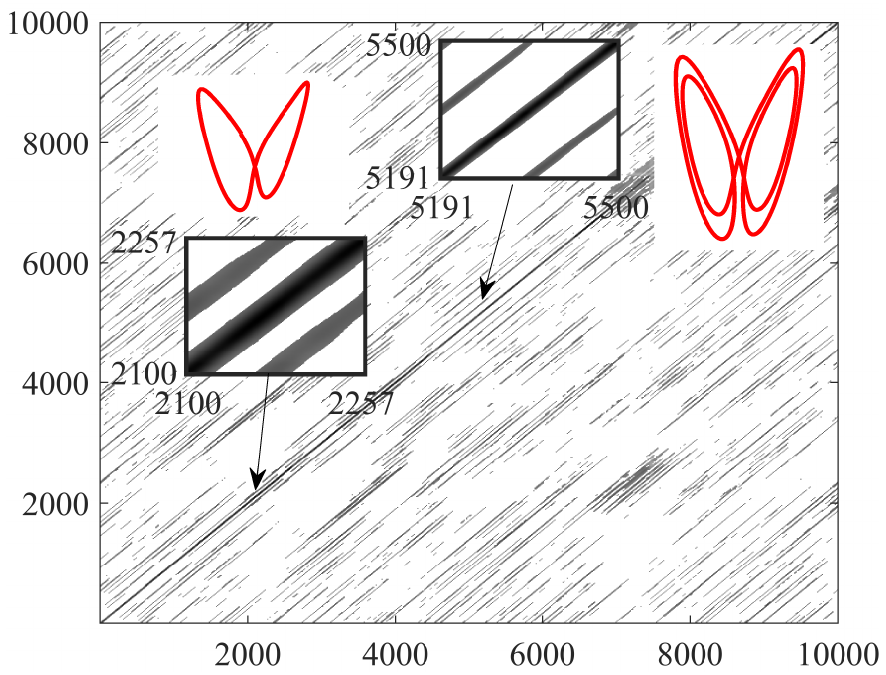}} 
    \subfloat[\label{}]{\includegraphics[width=0.49\columnwidth]{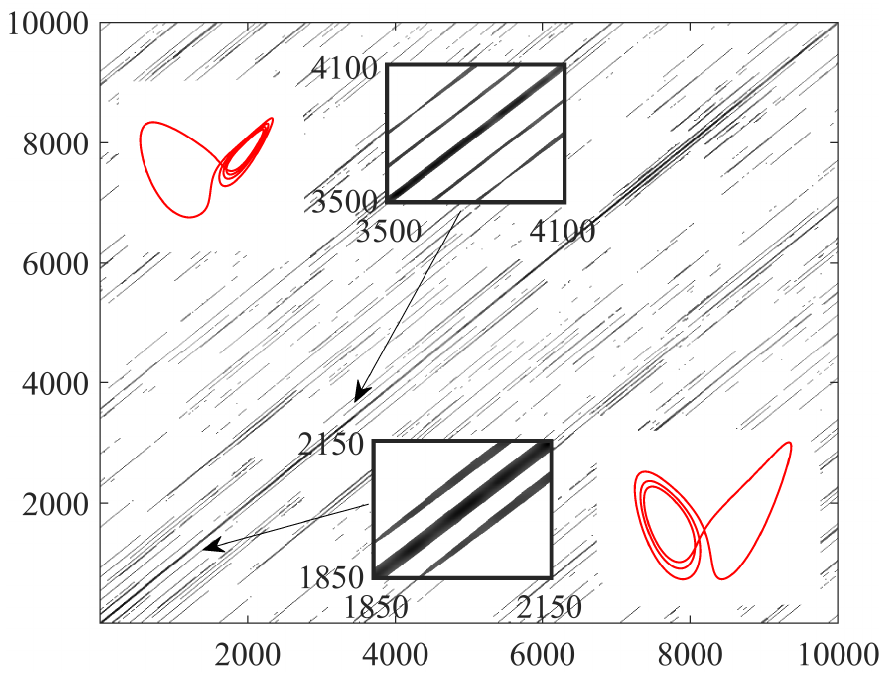}} 
    \caption{Detecting UPOs in Lorenz attractor by examining the Markov kernel operator $P_Q$. (a) Q=1000. (b) Q=2000.}
    \label{fig:kernel}
\end{figure}

Figure \ref{fig:kernel} illustrates examples of detected UPOs for the Lorenz attractor by examining the Markov kernel $P_Q$ and searching for regular patches (Figure \ref{fig:GA}) near different diagonal elements. We ran the Lorenz model from a random initial condition and sampled $N=10^4$ data points at a sampling rate $\Delta t=10^{-2}$s. The data was collected along a single long trajectory, allowing for a spin-up time to ensure the trajectory had settled onto the attractor before data collection. To reduce the computational burden for large sample sizes $N$, we sparsify $P_Q$ by selecting a cutoff value $k_{nn}\ll N$ and setting all but the largest $k_{nn}$ elements in each row of $K_Q$ to zero and symmetrizing the resulting sparse matrix. For simulation we set $k_{nn}=0.1N$. 

In Figure \ref{fig:kernel}, each regular patch indicates a periodic behavior for the corresponding time interval. We examine different values for the delay horizon, and Figure \ref{fig:kernel} shows the results for two values: $Q=1000, 2000$. According to Figure \ref{fig:kernel}, the delay horizon $Q$ allows for the recovery of UPOs at different scales. In this setting, increasing the delay horizon recovers UPOs with longer periods. In the limit of a large delay horizon (as $Q\rightarrow\infty$), the entire attractor is recovered as a UPO with a significantly long period.      

Recently, UPOs similar to those displayed in Figure \ref{fig:kernel} were identified using a variational approach \cite{Dong2020LorenzUPO}. Unlike that approach, we operate in a purely data-driven setting and use require no prior information on model equations. This is the advantage of diffusion maps \cite{Coifman2006Diffusion}, enhanced by delay-coordinate spaces, which boost their ability to extract intrinsic dynamical features and time scales.

\begin{figure}
      \centering
      \includegraphics[width=0.5\columnwidth]{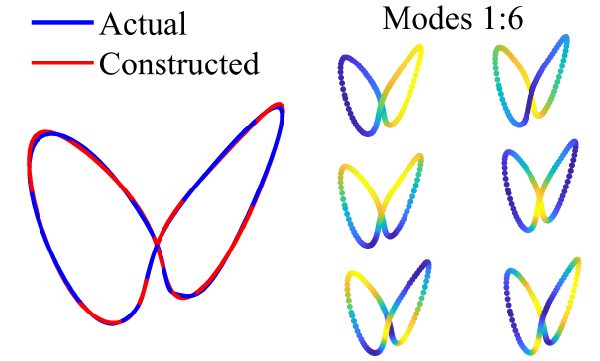}
      \caption{The periodic orbit reconstructed by 14 Koopman modes. Here the first six modes with least Dirichlet energy are displayed.}
      \label{fig:modes}
\end{figure}

We take further advantage of this approach in the eigenfunctions space and identify each detected UPO based on principle Koopman modes. 
For example, Figure \ref{fig:modes} shows the reconstruction of the UPO in the left of Figure \ref{fig:kernal_a}. It shows the first 6 Koopman eigenfunctions with the least Dirichlet energy values among the 14 Koopman eigenfunctions reconstructing the UPO. We ran the Lorenz model on detected UPO and sampled $N_s=1.4\times 10^3$ data points with a sampling rate of $\Delta t=10^{-2}$s. The Koopman eigenfunctions on the UPO were computed by the Galerkin approximation using kernel operator eigenfunctions as basis. This was achieved by a delay horizon that was twice the trajectory length along the UPO.  

The Koopman eigenfunctions set provides a multiscale family of geometric representations of the data, corresponding to dynamical features at different scales. From the random walk point of view of Markov processes \cite{Coifman2006Diffusion}, each eigenfunction scales with the probability of escaping different regions in state space. 
In this manner, the first eigenfunction in Figure \ref{fig:modes} is associated with recurrent transitions between the right and left UPO's lobes. The second eigenfunction reveals another bipartite splitting of the UPO. After the trivial constant Koopman eigenfunction, representing no transition over the UPO, the first two (nontrivial) eigenfunctions in Figure \ref{fig:modes} pertain to the slowest transition between different sets. 
Put another way, the probability of transitioning between the two yellow and blue sets in the first row of \ref{fig:modes} is the lowest within the Markov process. The third eigenfunction on the left of second row in Figure \ref{fig:modes}, indicates the next slow transition. It shows a slowly evolving trajectory at the bottom of the UPO (the lower blue part in the third eigenfunction). The trajectory reaching the lower section of the UPO undergoes a slow evolution and needs a longer time to escape this region. Therefore, slow transitions between the right and left UPO's lobes or between the upper and lower parts of the UPO are captured by principle Koopman modes. Subsequent eigenfunctions discover faster transitions.  

Therefore, detecting UPOs in the operator-theoretic framework benefits from an interpretable, data-driven setting that identifies the dynamics by decomposing them into the most basic mechanisms, or coherent patterns. This approach has initiated a new chapter in studying chaotic systems in the presence of complexities and challenges such as convergence of nearby trajectories or continuous spectra \cite{Mezic2005Spectral, Mezic2020Spectral, Giannakis2019Spectral, Colbrook2023resDMD}.   

 
\subsection{Controlling UPOs}\label{sec:LorenzCntrl}

We consider the nonlinear system as,
\begin{equation}
    \begin{gathered}
    \dot{\boldsymbol x}=F(\boldsymbol x,\boldsymbol u)    
    \end{gathered}
\end{equation}
with $\boldsymbol x\in\mathbb R^d$, $\boldsymbol y\in\mathbb R^m$, and $f:\mathbb R^{dm}\mapsto\mathbb R^d$ representing the system state, control input, and nonlinear map, respectively.
We consider a controlled Lorenz model as $F(\boldsymbol x,\boldsymbol u)=f(\boldsymbol x)+\boldsymbol u$, with $\boldsymbol u=[u_1 \ u_2 \ u_3]^T$ the control vector being applied linearly at each state direction. 

\begin{figure}
      \centering
      \includegraphics[width=1\columnwidth]{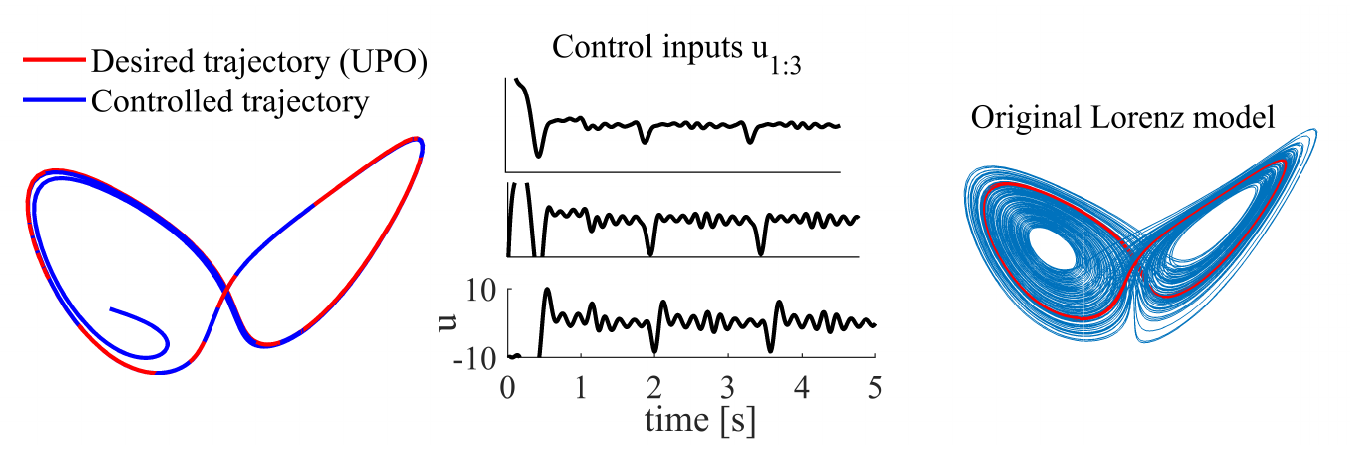}
      \caption{The controlled Lorenz attractor: tracking the UPO.}
      \label{fig:cntrl1}
\end{figure}

Our first control architecture utilizes the UPO's Koopman model \eqref{eq:x_d} as a reference trajectory. Detecting UPOs within strange attractors, and maintaining bounded trajectories in an ergodic setting enable us to achieve practical stabilization of UPOs throughlinear control actions \cite{Ott1990OGY, Ott_Book}. In the setting of this paper, this results in a trajectory tracking control law involving the proportional error feedback $\boldsymbol u=-K_u\boldsymbol e(t)$ where $K_u\in\mathbb R^{d\times m}$. The effectiveness  of this approach when applied to the Lorenz model, is illustrated in Figure \ref{fig:cntrl1}. It shows that, for $K_u=5I_{3\times 3}$, the initial chaotic trajectory converges to the selected UPO shortly after simulation begins. 

Figure \ref{fig:cntrl1} shows instances of larger control actions. These instances are characterized by more efforts needed to adhere to faster-evolving trajectories. Specifically, at the upper segment of the Lorenz attractor, substantial control actions are required to follow the selected UPO. 

In our second control architecture, to guarantee stability with minimal control action, the trajectories are allowed to evolve freely until they approach sufficiently close to the UPO (which happens infinitely many times in a chaotic attractor). 
At that point, control is strategically activated when the trajectory is adequately close to the UPO's stable manifold \cite{Ott1990OGY}.
The challenge lies in detecting the stable manifold in cases where a precise dynamical model is lacking or the system is not effectively low-dimensional. 

Identifying the evolving dynamics of UPOs using the approach developed in this paper offers an opportunity to empirically realize the positioning of trajectories near the UPO with respect to the stable manifold. It is important to note that not all trajectory near the UPO are accurately reconstructed by the UPO's Koopman eigenfunctions. Only trajectories dwelling near the UPO for a significant time can be reconstructed. More precisely, when considering dynamics adapted kernel operators in delay coordinates (see \ref{sec:Kernel}), the Nystrum extension of Koopman eigenfunctions is restricted to trajectory points that maintain a reliable trace along the UPO. Indeed, trajectories near the stable manifold are attracted to and  remain close to the UPO before being repelled by the unstable manifold. These  trajectories inherit the dynamical properties of the UPO, and the eigenfunctions extend to them reduced error.

\begin{figure}
      \centering
      \includegraphics[width=1\columnwidth]{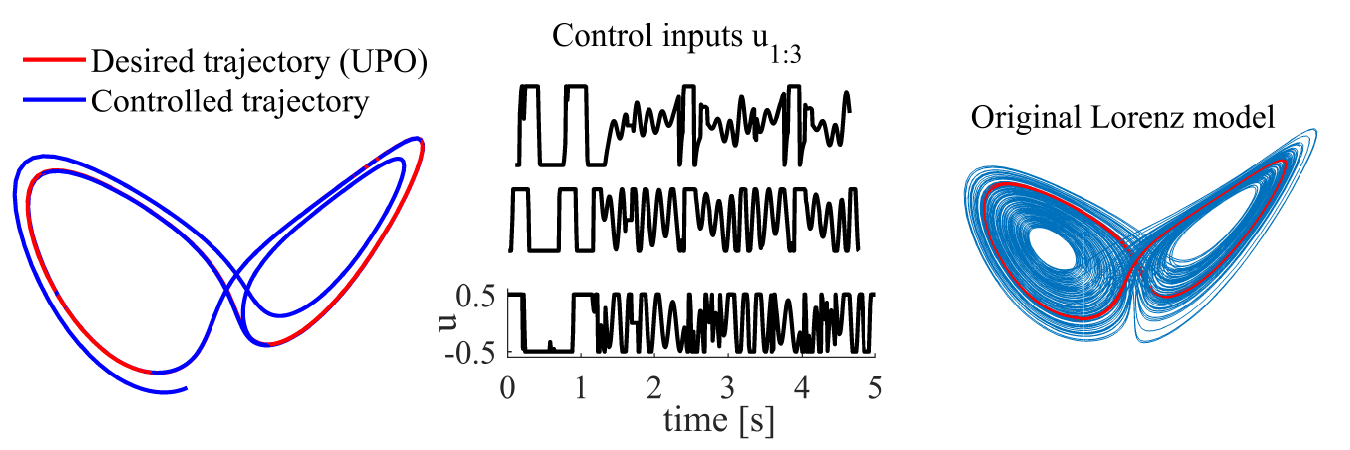}
      \caption{The controlled Lorenz attractor: small control-perturbation.}
      \label{fig:cntrl2}
\end{figure}

Therefore, we replace the physical distance (Euclidean distance in state space) with a measure in the UPO's eigenfunctions space. Crucially, we only trigger the control if the Koopman eigenfunctions of the UPO extend to the evolving trajectory with minimal error. Equation \eqref{eq:x_d} is used to assess this error based on the Koopman eigenfunction set $\hat{\boldsymbol\psi}(\boldsymbol x)=[\hat\psi_1(\boldsymbol x),\dots,\hat\psi_N(\boldsymbol x)]^T$, extending to the point $\boldsymbol x$ in the vicinity of UPO. 
The control law is defined based on the error $\boldsymbol e=\boldsymbol x-\hat{\boldsymbol x}$, with the reconstructed state $\hat {\boldsymbol x}=C\hat{\boldsymbol\psi}(\boldsymbol x)$, and the distance $d_x$ of $\boldsymbol x$ from the UPO, as,
\begin{equation}
    \begin{gathered}
        \boldsymbol u=\begin{cases}
            -K_ud_x,\hspace{5mm} \text{if}\hspace{1mm} \boldsymbol e<\varepsilon \\
            0,\hspace{10mm} \text{otherwise}
        \end{cases}
    \end{gathered}
\end{equation}
where $K_u$ is the control matrix gain and $\varepsilon$ a small positive scalar. See Figure \ref{fig:cntrl2} for the results of this approach applied to the Lorenz attractor. Here, we have restricted the control vector to satisfy $-0.5<u_i<0.5$, $i=1,2,3$. Figure \ref{fig:cntrl2} indicates the convergence of the chaotic trajectory to the UPO with small control perturbation. Compared to the trajectory shown in \ref{fig:cntrl1}, the trajectory in Figure \ref{fig:cntrl2} takes a longer time to settle onto the UPO, but the control action in \ref{fig:cntrl2} is significantly smaller. Thus, detecting and utilizing the ergodic properties in operator settings offer an effective, practical approach to tame chaos.



\section{CONCLUSIONS}\label{sec:cnclsn}

This paper demonstrates the use of operator settings to detect, identify, and stabilize UPOs in strange attractors. Delay embedding of the complex dynamical data into diffusion map coordinates offers a flexible approach for extracting UPOs across different time scales. The detection of UPOs is followed by a Koopman operator approach, identifing the principle dynamical modes on individual UPOs. 
This provides a more straightforward, interpretable tool to study complex strange attractors, based on Koopman eigenfunctions of different UPOs that are densely embedded within the chaotic attractor. It also introduces a unique method for stabilizing chaotic attractors using individual UPOs eigenfunctions. 
An extended version of this work will delve into more features of strange attractors in data-driven operator settings and their role in constructing more robust predictive and control models. Exploring the possible connection between the continuous spectra and pseudospectra \cite{Mezic2020Spectral, Giannakis2021RKHS,Colbrook2023resDMD} of strange attractors and different UPOs' spectra is particularly a compelling subject. 









\bibliographystyle{IEEEtran}
\bibliography{bibtex/bib/IEEEabrv.bib, bibtex/bib/refs.bib}{}

\end{document}